\documentclass[aps,prl,amsmath,amssymb,twocolumn,nofootinbib]{revtex4}
\usepackage{pslatex,graphicx,dcolumn,bm,natbib}
\usepackage{subfigure}
\date{\today}

\begin{document}

\title{Statistical Properties of a 2D Granular Material Subjected to
  Cyclic Shear}

\author{J. Zhang, T. S. Majmudar, A. Tordesillas$^2$ and R.P. Behringer$^1$\\
\normalsize{$^1$Department of Physics, Duke University, Box 90305,
Durham, NC 27708, USA\\$^2$Department of Mathematics and Statistics, University of Melbourne, Victoria 3010,
Australia} }

\begin{abstract}
This work focuses on the evolution of structure and stress for an
experimental system of 2D photoelastic particles that is subjected to
multiple cycles of pure shear.  Throughout this process, we determine
the contact network and the contact forces using particle tracking and
photoelastic techniques.  These data yield the fabric and stress
tensors and the distributions of contact forces in the normal and
tangential directions.  We then find that there is, to a reasonable
approximation, a functional relation between the system pressure, $P$,
and the mean contact number, $Z$.  This relationship applies to the
shear stress $\tau$, except for the strains in the immediate vicinity
of the contact network reversal.  By contrast, quantities such as $P$,
$\tau$ and $Z$ are strongly hysteretic functions of the strain,
$\epsilon$.  We find that the distributions of normal and tangential
forces, when expressed in terms of the appropriate means, are
essentially independent of strain.  We close by analyzing a subset of
shear data in terms of strong and weak force networks.
\end{abstract}

-\pacs{83.80.Fg,45.70.-n,64.60.-i}
\maketitle

\section{Introduction}

\begin{figure}
\centerline{\includegraphics[width=3in]{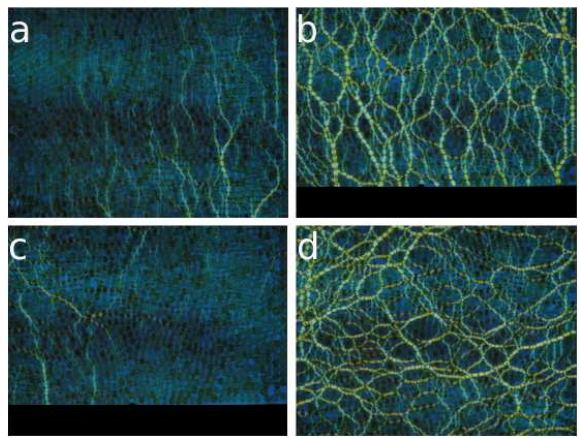}}
\caption{\label{fig:force-chains} Sequence of photoelastic images
  showing the evolution of the force chains as the system is sheared
  in the forward (images a and b), then the reverse direction (image c
  and d). These four images are chosen at different steps from the 1st
  shear cycle, with $\epsilon=0.033$, $0.267$, $0.267$, and $0.033$,
  respectively.  The axial strain $\epsilon$ is defined below.  In
  images b and c, the sidewall of the biax has moved into view,
  creating dark bands at the bottom of the images.}
\end{figure}

In this work, we describe experiments that probe the microscopic
properties of sheared granular materials, with an eye towards
understanding the statistical properties and small-scale phenomena
which strongly influence larger scale behavior.  The application of
shear leads to the evolution of a strong force network, as shown in
Fig.~\ref{fig:force-chains}, sometimes referred to as force chains.
These mesoscopic structures are filamentary networks that carry forces
at or above the mean, and that extend, in the case of shear, over
distances of a few to perhaps many tens of grains.  During shear, the
force network evolves, with force chains strengthening, and then
ultimately failing.

The present studies explore the structural evolution of system during
shear by means of fabric, stress and related tensors.  Associated with
the evolution of the fabric and stress tensors are a number of complex
phenomena, including shear bands, particle rotation, failure and
buckling of the force chains, among other effects. We have recently
shown the importance of rotation for the failure of force chains,
particularly in shear bands \cite{tordZhangBehr}. In general, all
quantities measured here show fluctuations, and, of course sensitivity
to the direction of the shear (forward and reverse). In particular,
when the shear direction is switched, the system undergoes structural
reconstruction, causing changes in the average contact number, the
mean orientation of the contacts, and the stress tensor.

We focus on a path corresponding to pure shear strain, starting from a
packing fraction where there is no observable stress.  As we strain
the system, the detected stresses and mean contact number $Z$
increase, and the system reaches a jammed state for $Z$'s above $Z
\simeq 3$.  As we further deform the system, including reversal of the
strain, $Z$ tends to remain at or above 3 for much of the time.
Throughout, multiple shear cycles, the packing fraction $\phi$ remains
at a fixed value, $\phi = 0.758$, that is below the observed jamming
value for isotropic compression\cite{majmudar07}.  When the shear
strain is reversed, the original force network largely vanishes, and a
new strong network forms.  This process is strongly hysteretic in the
strain, but we find that the stresses can be characterized rather well
in terms of the system-averaged contact number, $Z$.

In the remainder of this work, we first describe basic features of the
experimental techniques.  We then present results from cyclic pure
shear experiments, in which we explore the structural and stress
changes within each shear cycle and in particular during shear
reversals. We then analyze the force network in terms of strong and
weak components.

\section{Experimental Techniques}

\begin{figure}
\centerline{\includegraphics[width=2.5in]{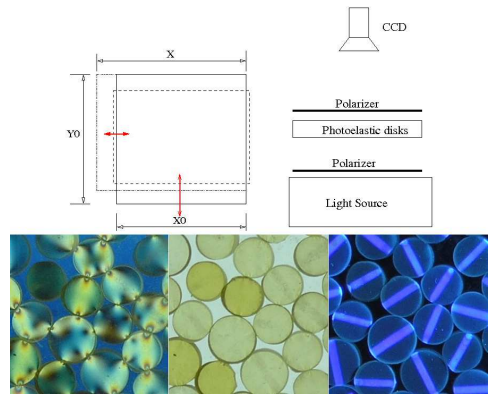}}
\caption{\label{fig:expt-images} Top-left: Sketch of top view of the
  experimental apparatus, a 2D `biax', consisting of pairs of facing
  boundaries that can be moved precisely under computer control so as
  to produce desired strains.  Particles rest on a smooth slippery
  sheet of Plexiglas and are confined laterally by the walls of the
  biax. Strains are applied quasi-statically, in small discrete
  steps. Top-right: Side view of apparatus.  Imaging is carried out by
  a camera mounted above the biax, and for each step, we obtain three
  images: one with crossed polarizers (bottom left), one without
  polarizers (bottom center), and one without polarizers but with UV
  illumination (bottom right).}
\end{figure}

The experiments described here use a `biaxial' apparatus, as depicted
in the upper parts of Fig.~\ref{fig:expt-images}.  This device allows
us to deform a rectangular sample of particles into any other desired
rectangular shape, hence apply pure shear (compression in one
direction, but equal dilation in the other), uniaxial compression,
isotropic compression, etc.  Here, we focus uniquely on shear
deformations, which maintain a fixed area for the system.

The studies are carried out effectively in 2D by using disks which are
made of photoelastic material. When under stress, and when viewed
between crossed polarizers, photoelastic materials exhibit a series of
light and dark bands, as in the bottom left image of
Fig.~\ref{fig:expt-images}. These bands encode the detailed stress
within each particle, and these stresses are in turn, determined by
the forces at contacts on each particle.  In the past, this technique
has been used in several different studies \cite{dantu,joselin-jong}.
What makes our current approach unique is that, for large collections
of particles, we solve the inverse problem which starts from the
photoelastic image and yields as output, the inter-particle contact
forces.  More details have been given elsewhere
\cite{majmudar05,tordZhangBehr,majmudar_thesis}.  The particles are
also typically marked with a small bar, which allows us to track the
rotation and displacement of individual particles.  In our current
tracking approach, the bars are drawn on with fluorescent ink, which
is invisible under ordinary light but glows strongly under UV light.
In this way, it is possible to have both photoelastic images for
force/stress measurement and separate images for tracking rotation and
displacement of a given set of particles, without mutual interference.
In earlier versions of this approach
\cite{howell99a,veje99,utter04a,utter04b,utter08}, we used solid black
bars drawn on the particles. In this case, we imaged separate sets of
particles for determining forces and for tracking motion.

The initial boundaries of the system form a square filled with 1568
bi-disperse photoelastic disks at a packing fraction $\phi = 0.758$.
There are roughly $80\%$ smaller particles having a diameter of
$0.74$~cm, and $20\%$ larger particles having a diameter of $0.86$~cm.
The initial state is prepared as close to isotropic as possible and is
stress-free. The system is then subjected to shear by compression
along the $y$-direction and expansion along the $x$-direction, as in
Fig.~\ref{fig:expt-images} (top-left), keeping the system area
constant. Since the area is fixed, the deformation can be simply
defined using the strain $\epsilon$ along the $x$-axis with
$\epsilon=(x-x_0)/x_0$. Here, $x_0$ is the initial size of the
square. Once a maximum deformation $\epsilon_{max}$ is reached, shear
is reversed by compression along the $x$-axis and expansion along the
$y$-axis. After this first shear reversal, the shear continues until
the system domain returns to a square, and then deforms in the
negative strain direction. Once $\epsilon$ reaches a minimum
$\epsilon_{min} < 0$, a second shear reversal is applied, eventually
returning the system domain to a square with $\epsilon=0$.  This
completes one shear cycle. The second shear cycle continues from the
final state of the first shear cycle and the same procedure is applied
for a total of six shear cycles. Note that the actual values of
$\epsilon_{max}$ and $\epsilon_{min}$ are different for each shear
cycle and the possible extreme values for these are determined by the
spatial limit of the apparatus. A list of $\epsilon_{min}$ and
$\epsilon_{max}$ are summarized in Table~\ref{epsilonlist}. The whole
shear process is carried out in small incremental quasi-static
steps. From one step to the next, $\epsilon$ increases or decreases by
a small amount $\delta\epsilon=\pm 3.3\times10^{-3}$, depending on the
shear direction. After each step, the motion is paused and images are
acquired.  The three images in the bottom row
Fig.~\ref{fig:expt-images} show close-ups of the three different image
types.  The left-most of these is taken with polarizers in place, the
middle is without polarizers and with ordinary light, and the
right-most is without polarizers and with UV light.

\begin{table}
\caption{\label{epsilonlist} A list of $\epsilon_{max}$ and $\epsilon_{min}$ for different shear cycles. }
\begin{tabular}{|l|l|l|}
\hline
Shear cycle & $\epsilon_{max}$ & $\epsilon_{min}$ \\
\hline
1 & 0.2867 & -0.15 \\
2 & 0.29 & -0.15 \\
3 & 0.2333 & -0.15\\
4 & 0.2 & -0.15 \\
5 & 0.1833 & -0.1667 \\
6 & 0.15 & N/A \\
\hline
\end{tabular}
\end{table}

\vspace{0.2in}

Before we turn to detailed results we note an experimental issue of
importance.  During the parts of the cycle where the overall stresses
in the system are low, the photoelastic response at some contacts
falls below our limit of resolution.  Because, as discussed below, the
experiments indicate a distribution of normal contact forces of the
form 
\begin{equation}
P(F_n) = \langle F_n \rangle^{-1} f(F_n/\langle F_n \rangle),
\end{equation}
where $f$ is to a reasonable approximation, the same function for all
mean forces, we can estimate the number of the missed contacts
reasonably well.  Non-zero contact forces below our experimental
resolution also affect our measurements of stress and, in particular,
$P$.  However, the effect on stress components is much lower, since
the contact forces appear linearly in the appropriate sums.

We expect that we miss a fraction 
\begin{equation}
\int_0^{F_c}
P(F_n)dF_n/\int_0^{\infty} P(F_n)dF_n
\end{equation}
of contacts, where $F_c$ is a small known cut-off force, roughly the
weight of a particle, below which we cannot detect the photoelastic
response.  This means that the measured $Z$'s are lower than their
true values by 
\begin{equation}
\int_{F_c}^{\infty} P(F_n)dF_n/\int_0^{\infty} P(F_n)dF_n.
\end{equation}
We also underestimate the pressure by a factor of
\begin{equation}
\int_{F_c}^{\infty} F_n  P(F_n)dF_n/\int_0^{\infty} F_n P(F_n)dF_n.
\end{equation}
In this last expression, we assume that all particles have the same
radius, which is a reasonably good assumption.  In this regard, the
correction can be as much as 15\% in $Z$ very near jamming, but then
becomes negligible for $Z$ a bit above 3.0.  The correction to $P$ is
much smaller, only 1-2\% close to the jamming transition.  In order to
simplify the correction, we assume that the force distribution is an
exponential.  This is roughly right, and produces a reasonable
correction, given the statistical variability of the data. The results
of $P$ and $Z$ presented in this paper have been corrected
accordingly.  We use the same correction factor for shear stresses,
$\tau$, although in this case, the correction is not a rigorous.

\section{Experimental Results}

\begin{figure}
\centerline{\includegraphics[width=2.5in]{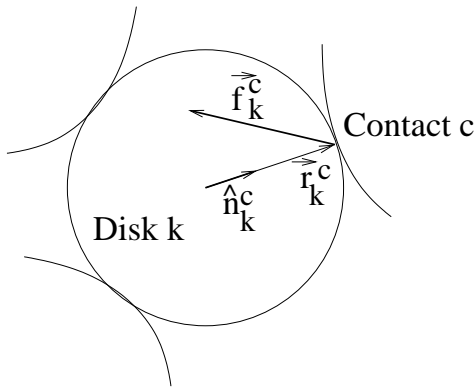}}
\caption{\label{fig:contact-sketch} Sketch illustrating the notation
  for calculating the fabric, force-moment and stress tensors.}
\end{figure}

We are concerned with the evolution of the force and contact networks.
Both are typically strongly anisotropic, and the direction of the
anisotropy switches quickly when the direction of strain is reversed.
The force anisotropy is evident in Fig.~\ref{fig:force-chains}, which
shows representative photoelastic images during different phases of a
single cycle.  The structural changes of the contact network during
cyclic shear are also strongly anisotropic.  These can be captured by
the fabric tensor, $R_{ij}$, defined as
\begin{equation}
R_{ij}=\frac{1}{N}\sum_{k=1}^N\sum_{c=1}^{c_k}n_{ik}^cn_{jk}^c. 
\end{equation}
Here, the summation and $N$ include only non-rattler disks, and as
illustrated in Fig.~\ref{fig:contact-sketch}, $c_k$ is the number of
contacts on disk $k$, and $n_{lk}^c$ is the $l$th component of the
unit branch vector pointing from the center of the disk $k$ to a
contact $c$. We consider a rattler disk to have less than two
detectable contacts.  The average contact number $Z$ is simply the
trace of the fabric tensor $R_{ij}$.  The principal eigendirection of
$R_{ij}$ is also a useful measure of the prevailing orientation of the
force network.

\begin{figure}
\centerline{\includegraphics[width=2.5in]{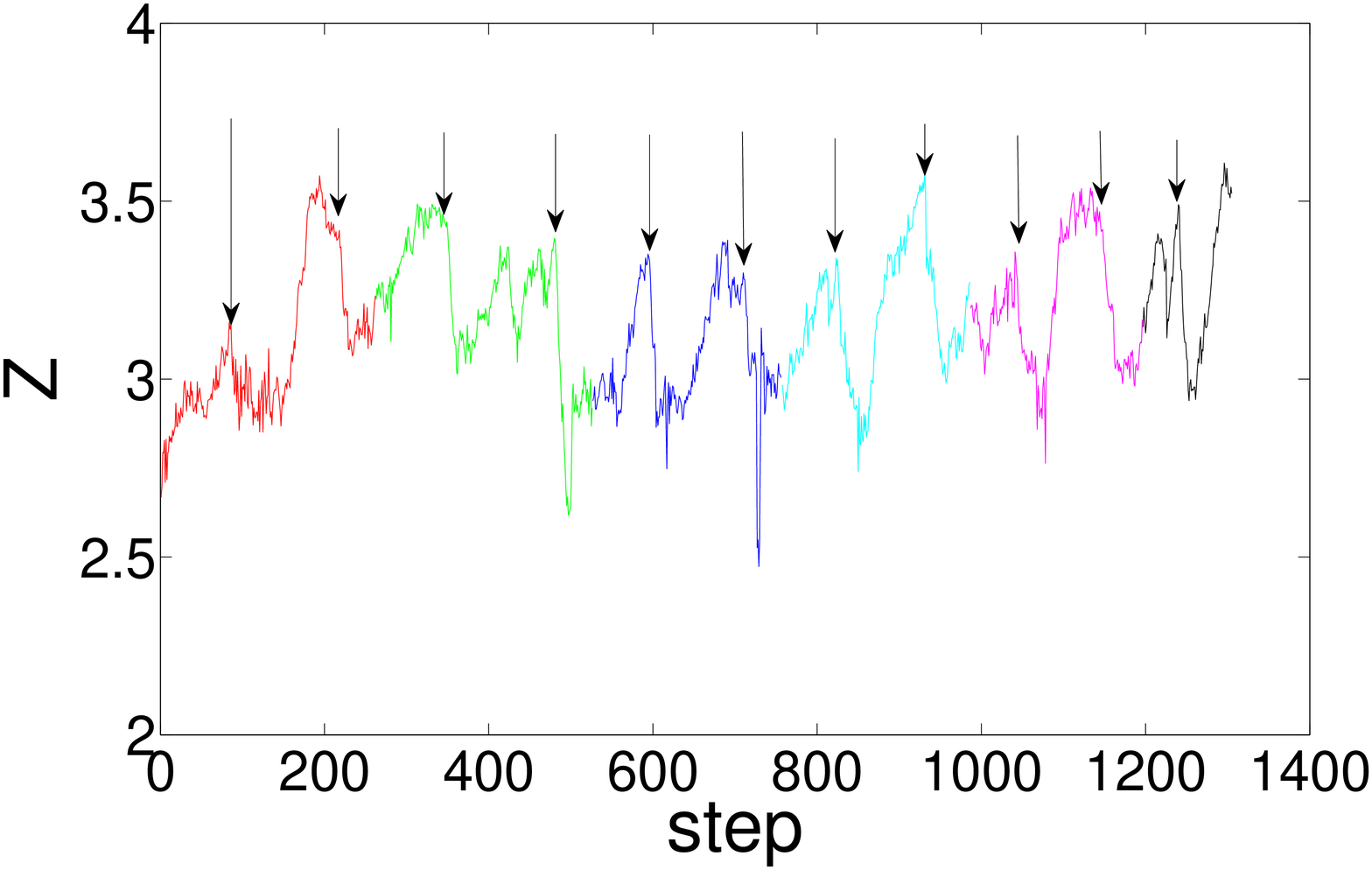}}
\caption{\label{fig:z_strain} Evolution of the average contact number
  $Z$ vs. step number $N_s$. Each shear cycle is colored
  differently. Arrows indicate the steps where the shear direction is
  switched.}
\end{figure}

\begin{figure}
\centerline{\includegraphics[width=2.5in]{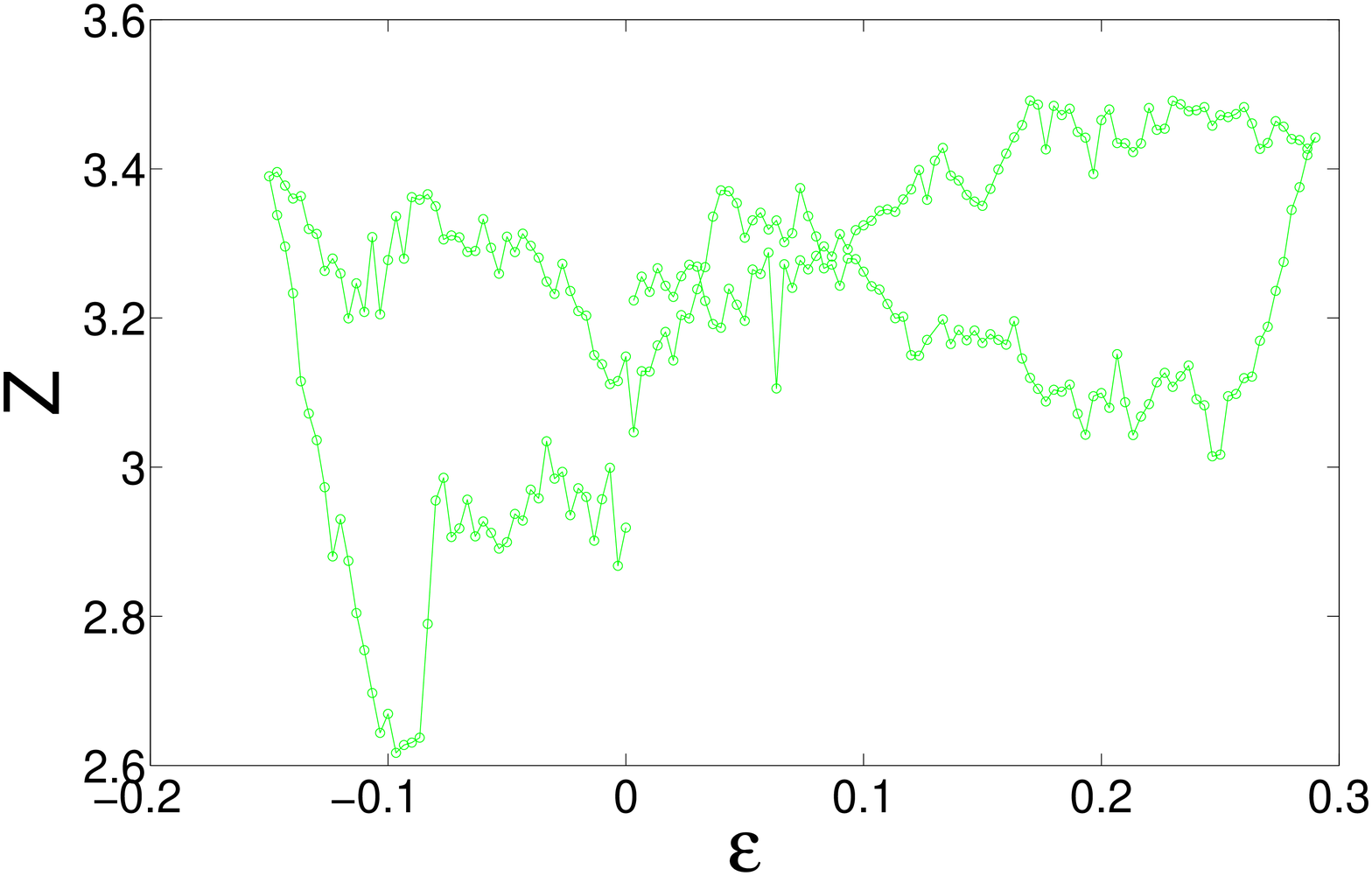}}
\caption{\label{fig:z_half} Mean contact number $Z$ vs. strain
  $\epsilon$ for the second cycle.}
\end{figure}

Figs.~\ref{fig:z_strain} and \ref{fig:z_half} show how $Z$ changes in
each shear cycle as a function of step number $N_s$ for all cycles,
and as a function of strain $\epsilon$ for one cycle.  In
Fig.~\ref{fig:z_strain}, the individual shear cycles are distinguished
using different colors.  We will maintain this color scheme throughout
to identify the various cycles.  Arrows in Fig.~\ref{fig:z_strain}
indicate the shear reversals.  $Z$ fluctuates between a minimum value
of around $2.5$ and a maximum value of around $3.5$.  The first shear
cycle, the red curve, begins with a nearly stress-free and isotropic
state.  The force chains build up steadily.  As a consequence, $Z$
increases as more force chains develop.  $Z$ barely exceeds 3 before
the first shear reversal. Immediately after the reversal, $Z$ drops.
The relatively rapid decrease of $Z$ after a switch of the shear
direction is common to all shear reversals because the force
network/force chains switch direction during this transition.  With
continued strain after a reversal, $Z$ again increases as a new strong
network, orthogonal to its predecessor, emerges.  Note that $Z=3$ is
the nominal isostatic point for frictional particles in two
dimensions.  Fig.~\ref{fig:z_strain} shows that the system remains
mostly in a jammed state after the first cycle, but that it may leave
a jammed state briefly after shear reversal.  As we show below, shear
bands form in response to the shear.  Hence, these states are not
spatially homogeneous.  However, it is the presence of a mechanically
rigid but dilated region in the shear bands that allows for jammed
states at $\phi$'s lower than the isotropic value.  We also emphasize
that $Z$ is strongly hysteretic when viewed with respect to strain.
To demonstrate this point, we show one shear cycle, the second, as a
function of strain in Fig.~\ref{fig:z_half}.

\begin{figure}
\centerline{\includegraphics[width=2.5in]{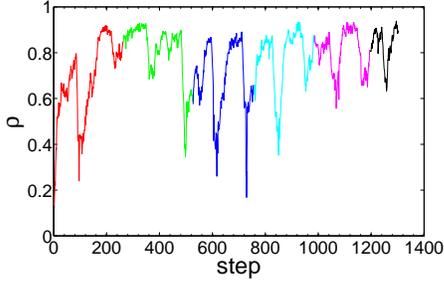}}
\caption{\label{fig:rho_steps} Evolution of the ratio, $\rho$,
  vs. step number, $N_s$. $\rho$ is defined as the number of
  non-rattler particles over the total number of particles.}
\end{figure}

At various steps, the fraction, $\rho$, of detectable non-rattler
particles also fluctuates, as displayed in
Fig.~\ref{fig:rho_steps}. This ratio changes from $0.2$ up to
$0.90$. Both Fig.~\ref{fig:z_strain} and Fig.~\ref{fig:rho_steps} show
similar trends, although $\rho$ is noisier at the step where shear is
reversed.

\begin{figure}
\centerline{\includegraphics[width=2.5in]{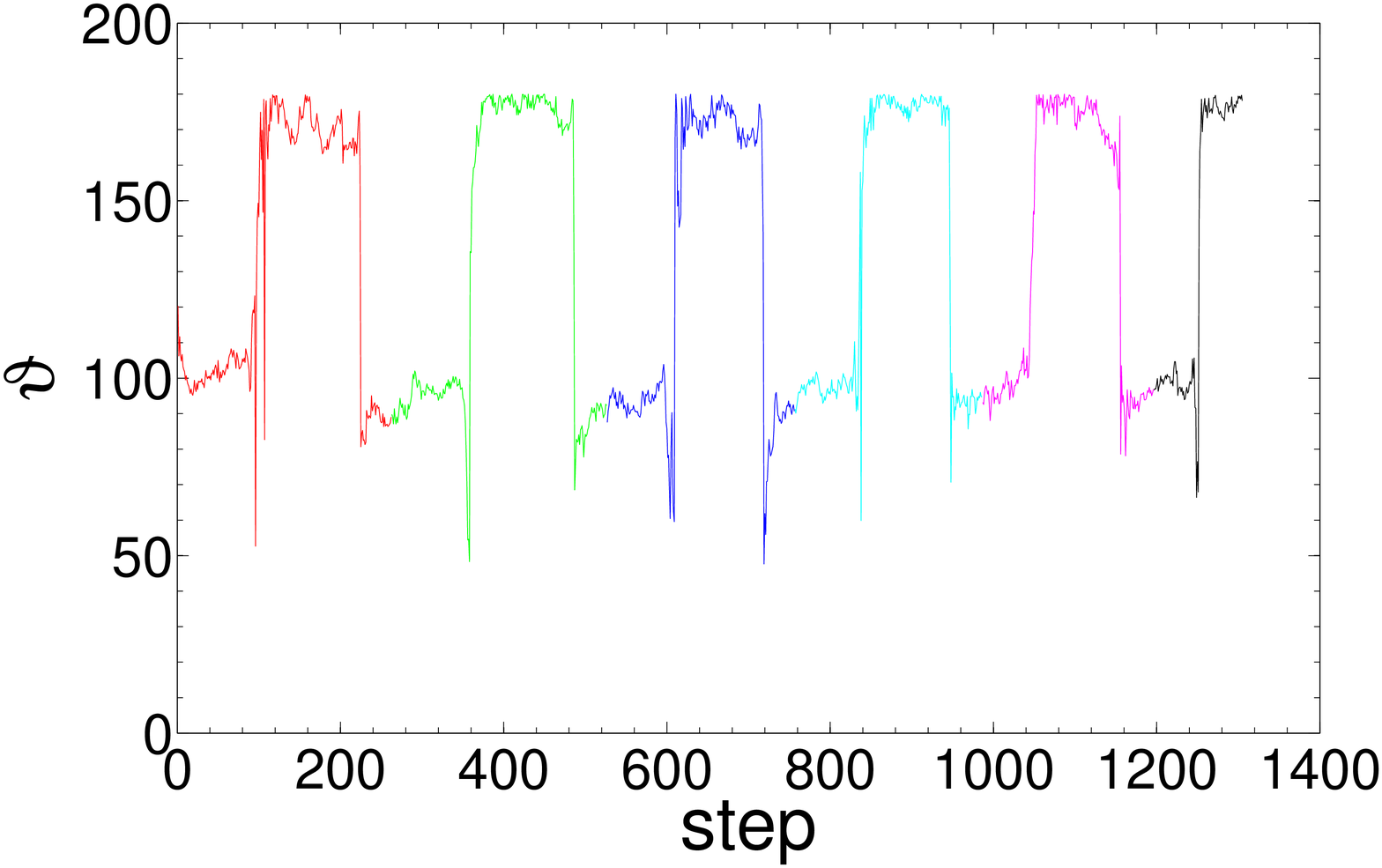}}
\caption{\label{fig:contactAngle_fw} Fabric orientation angle
  $\vartheta$ vs.  step number, $N_s$. $\vartheta$ is defined as the
  absolute angle between the eigenvector of the maximum eigenvalue of
  the fabric tensor $R$ and the $x$ axis. This angle measures the
  dominant contact orientation}
\end{figure}
\begin{figure}
\centerline{\includegraphics[width=3.5in]{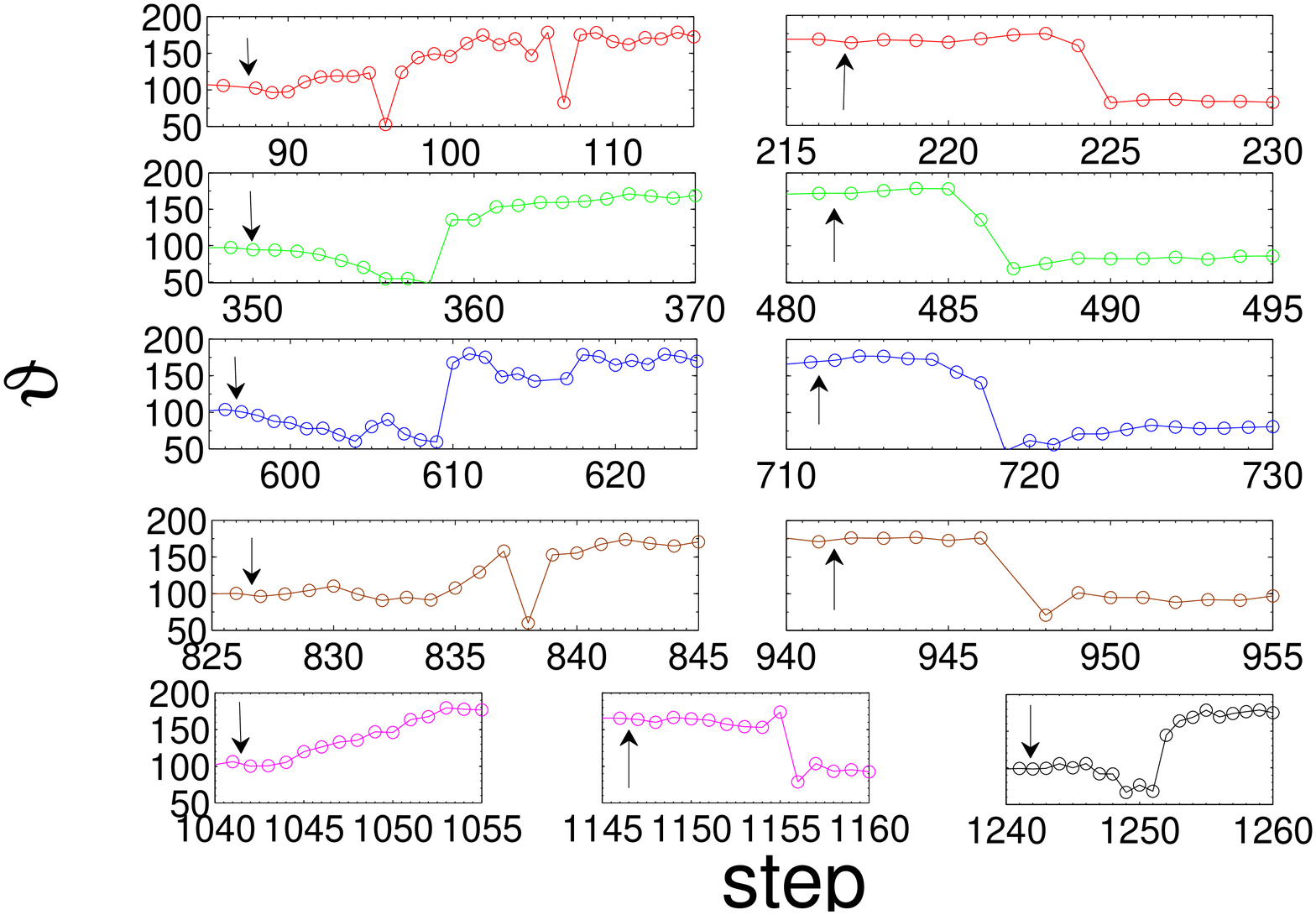}}
\caption{\label{fig:contactAngle_finer} Fabric orientation angle
  $\vartheta$ vs. step number $N_s$ on a fine scale near shear
  reversals.  Arrows in each graph indicate the beginning of the shear
  reversals. The readjustment of $\vartheta$ after a shear reversal
  takes about 10 to 20 steps.}
\end{figure}

We characterize the mean anisotropy of the contact network in terms of
$\vartheta$, the system-averaged value of the angle between the
eigenvector of the maximum eigenvalue of $R$ and the $x$ axis.  Here,
we restrict $0 < \vartheta \leq 180^o$.
Figure~\ref{fig:contactAngle_fw} shows that $\vartheta$ switches
quickly, shortly after each strain reversal.  That is, after a very
small strain, $\vartheta$ aligns with the compressive direction.  In
order to see how quickly the angle changes after a shear reversal, we
have plotted $\vartheta$ on a much finer scale. The results are
presented in Fig.~\ref{fig:contactAngle_finer}, where the graphs are
organized from top to bottom, as a function of step number, $N_s$. The
typical number of steps required for the readjustment of the
orientation varies from $\Delta N_s \simeq 10$.  However, we note that
there is often a lag between the strain reversal, and the switch in
$\vartheta$, which may occur in only a few strain steps.

\begin{figure}
\centerline{\includegraphics[width=2.5in]{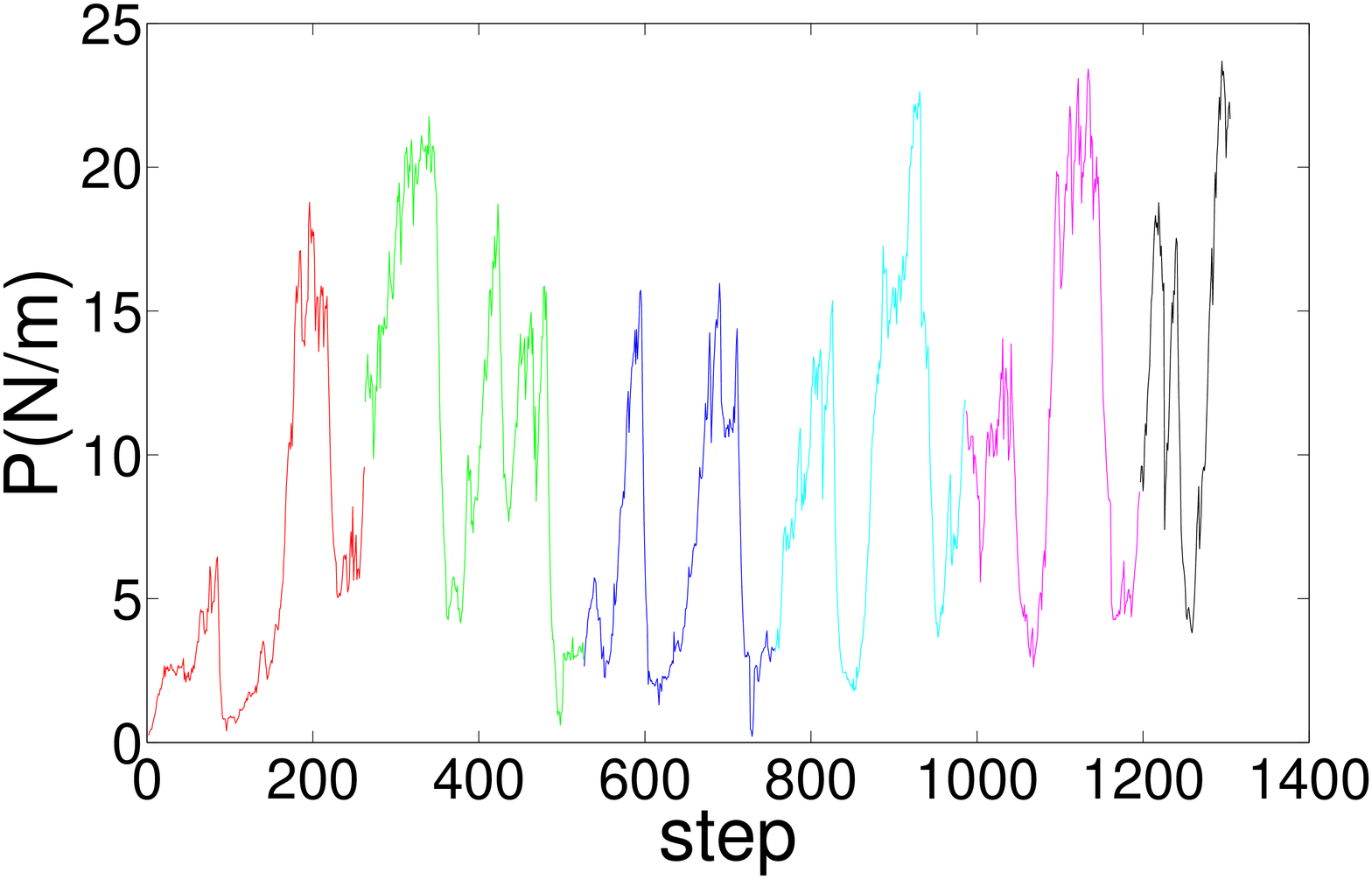}}
\caption{\label{fig:pressure_step} Pressure $P$ vs. step number
  $N_s$.}
\end{figure}

\begin{figure}
\centerline{\includegraphics[width=2.5in]{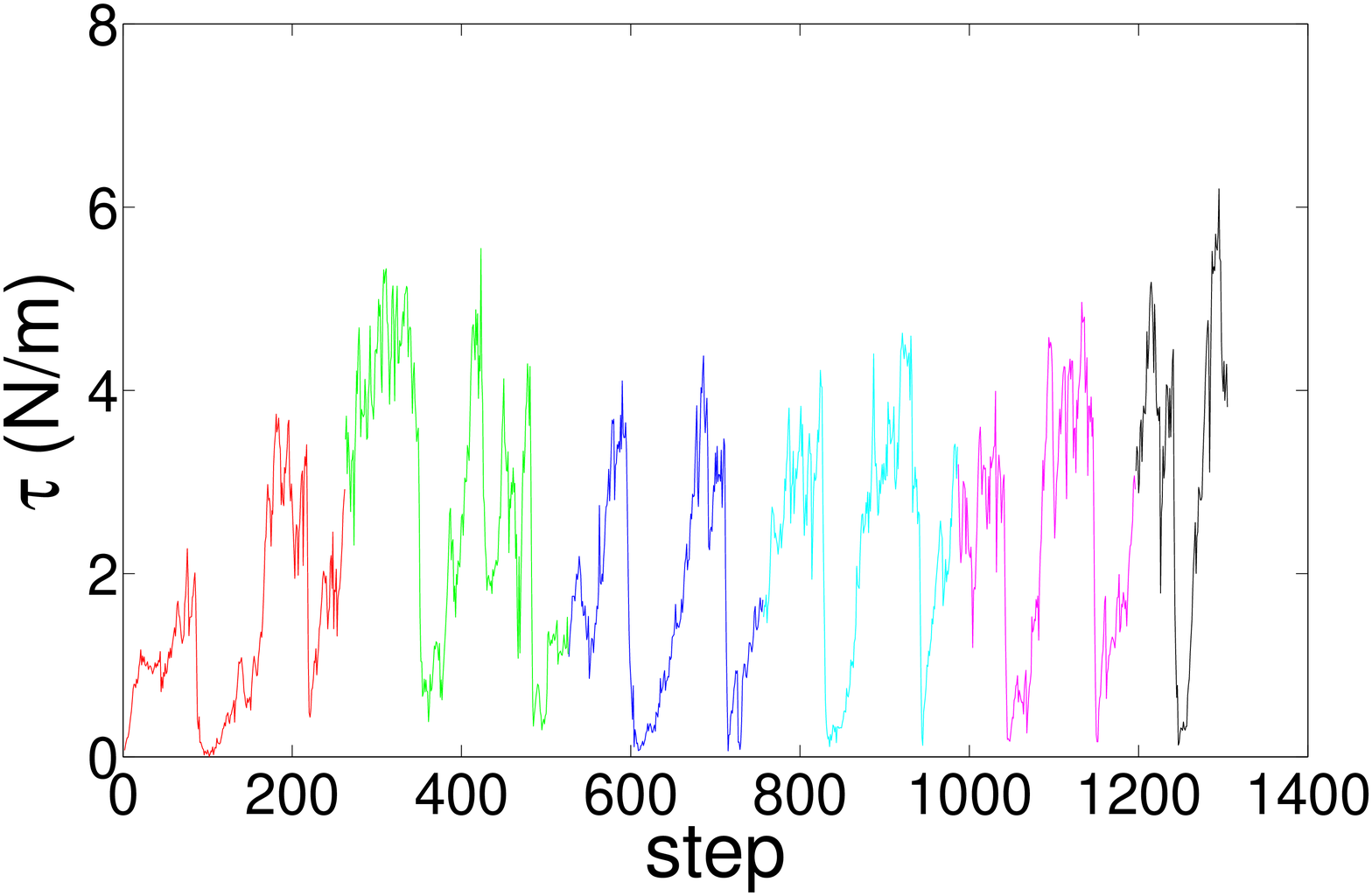}}
\caption{\label{fig:shearstress_step} Shear stress $\tau$ vs. step
  number $N_s$.}
\end{figure}

\begin{figure}
\centerline{\includegraphics[width=2.5in]{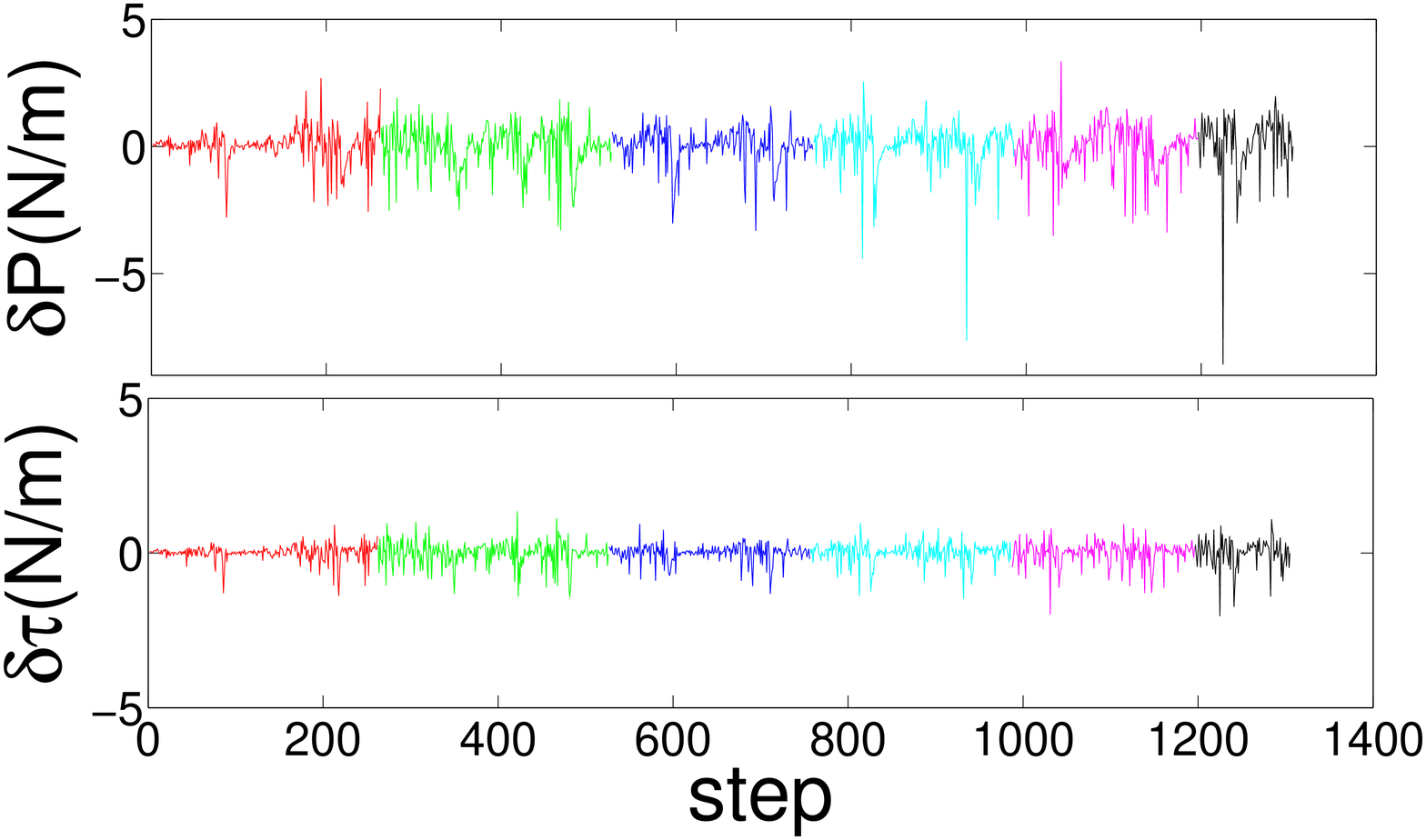}}
\caption{\label{fig:dpdausteps} $\delta P$ (top) and $\delta \tau$
  (bottom) vs. step number $N_s$. $\delta P$ and $\delta \tau$ are
  respective differences between two neighboring steps of $P$ and
  $\tau$ as given in Fig.~\ref{fig:pressure_step} and
  Fig.~\ref{fig:shearstress_step}.}
\end{figure}

\begin{figure}
\centerline{\includegraphics[width=2.5in]{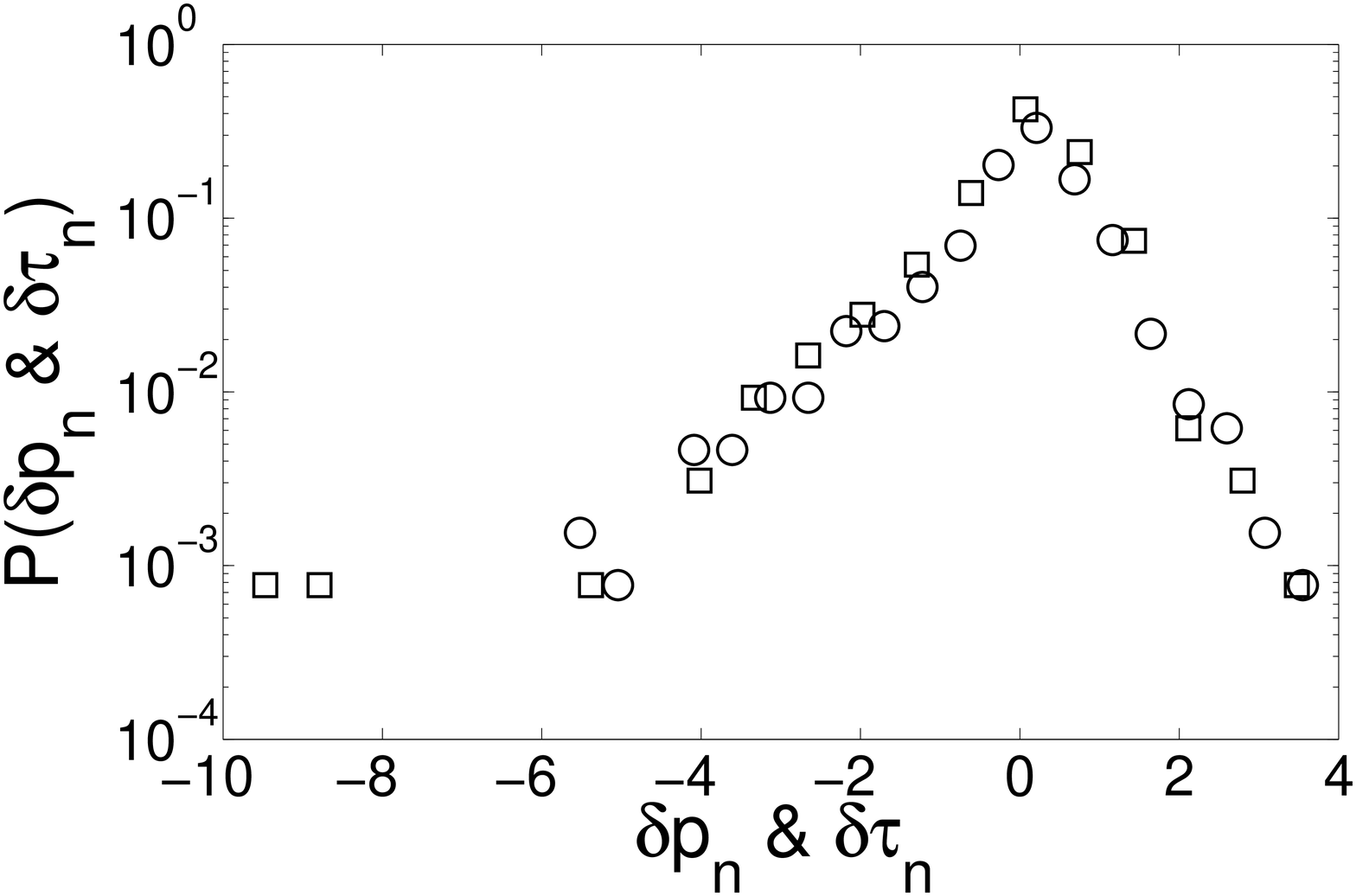}}
\caption{\label{fig:dpdtauhist} PDFs of $\delta p_n$ (circles), and
  $\delta \tau_n$ (squares). $\delta p_n$ and $\delta \tau_n$ are made
  dimensionless through normalization by their corresponding standard
  deviations.}
\end{figure}

\begin{figure}
\centerline{\includegraphics[width=2.5in]{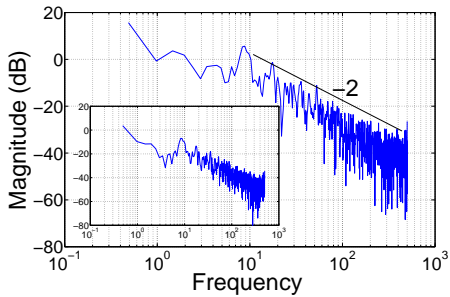}}
\caption{\label{fig:ptauspectrum} Power spectra, on log-log scales,
  for $\delta p_n$, (inset) and $\delta \tau_n$ computed from the data
  given in Fig.~\ref{fig:pressure_step}
  Fig.~\ref{fig:shearstress_step} vs. frequency corresponding to
  inverse step number, where a frequency of $1.0$ corresponds to a
  $1/1000$~steps.  The solid line is a guide to the eye, and
  corresponds to a power law with an exponent of $-2$.}
\end{figure}

The stress tensor $\sigma_{ij}$ and the force moment tensor,
$\hat{\sigma}_{ij}$ provide additional measures of anisotropy, in this
case for the forces.  We define a local force moment tensor as
\begin{equation}
\hat{\sigma}_{ij}= \sum_{c=1}^{c_k}f_{ik}^cr_{jk}^c.
\end{equation}
The globally averaged stress tensor is then
\begin{equation}
\sigma_{ij}=\frac{1}{A}\sum_{k=1}^N\hat{\sigma}_{ij}.
\end{equation}
Here, $A$ is the system area; $N$, $c_k$, $i$, $r_{jk}^c$ and $j$ have
the same meaning as in the expression of $R_{ij}$
(e.g. Fig.~\ref{fig:contact-sketch}). $f_{ik}^c$ is the $i$th
component of the contact force on particle $k$ at contact $c$.  The
two eigenvalues of the stress tensor are $\sigma_1$ and $\sigma_2$,
where $\sigma_1 \leq \sigma_2$ by definition. The pressure is then
$P=\frac{1}{2}(\sigma_1+\sigma_2)$ and the shear stress is
$\tau=\frac{1}{2}(\sigma_2-\sigma_1)$.

\begin{figure}
\centerline{\includegraphics[width=2.5in]{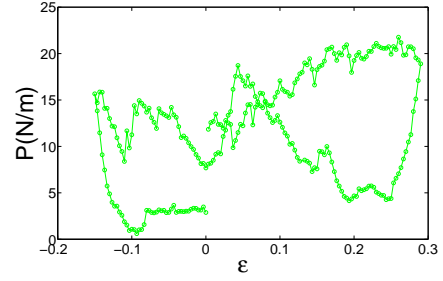}}
\caption{\label{fig:p_half} Pressure, $P$ vs. strain, $\epsilon$, for
  the second cycle, showing strong hysteresis.}
\end{figure}

\begin{figure}
\centerline{\includegraphics[width=2.5in]{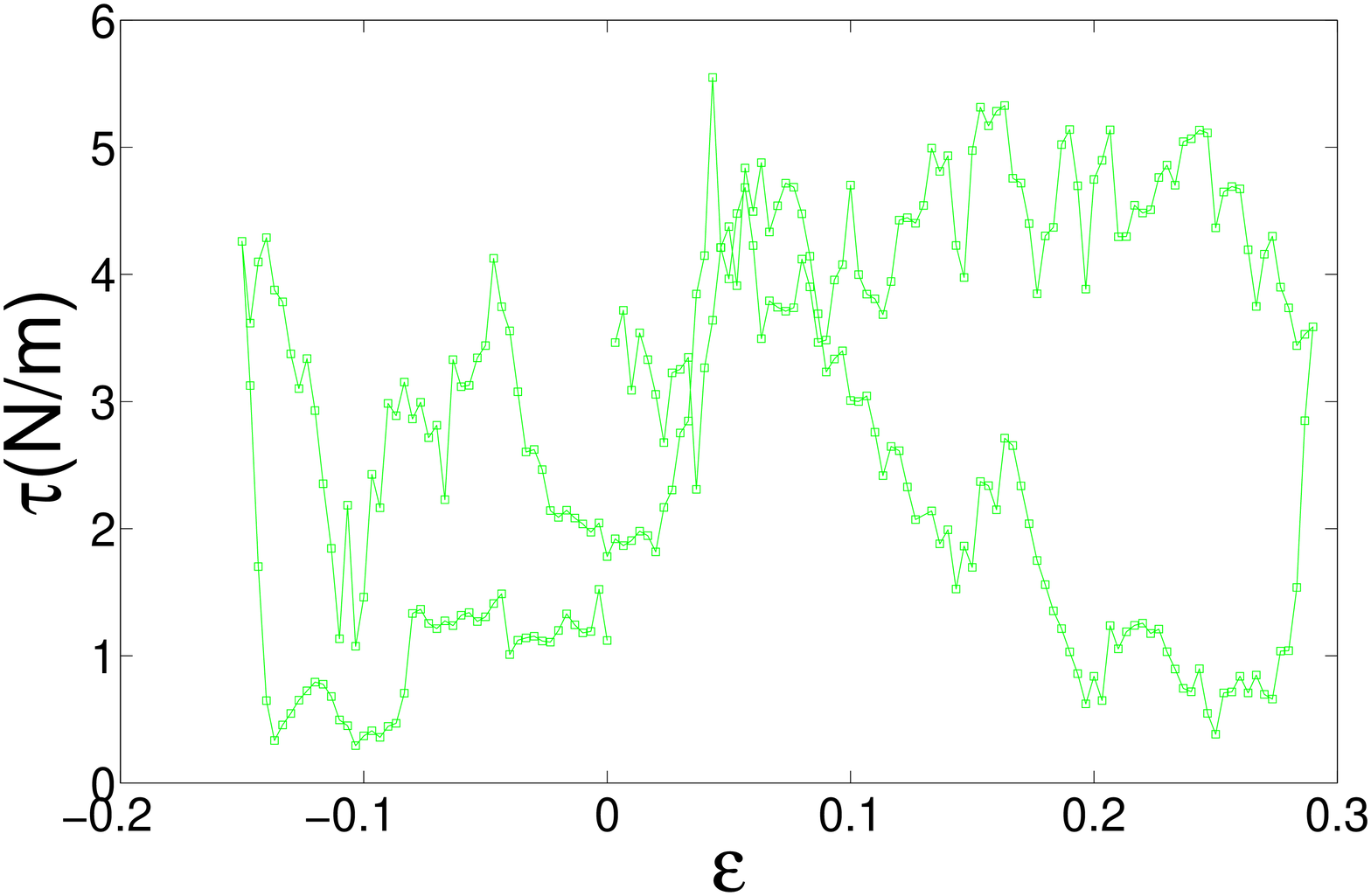}}
\caption{\label{fig:tau_half}Shear stress $\tau$ vs. strain,
  $\epsilon$ for the second cycle.}
\end{figure}

{Figure~\ref{fig:pressure_step} and Fig.~\ref{fig:shearstress_step}
  show the evolution of the pressure and the shear stress versus step
  number, $N_s$, and Figs.~\ref{fig:p_half} and \ref{fig:tau_half}
  show $P$ and $\tau$ over the second cycle vs. strain.  Both $P$ and
  $\tau$ } vary significantly with strain, but unlike $\vartheta$,
both quantities evolve steadily, modulo some significant fluctuations,
up to the maximum of a given cycle.  The fluctuations evident in the
stress evolution are due to failure events, which are associated with
a continual collapse of old and formation of new force chains. To show
the scale of the fluctuations more clearly, we present in
Fig.~\ref{fig:dpdausteps}, the changes $\delta P$ and $\delta \tau$ in
$P$ and $\tau$, respectively, between successive steps.  The curves
exhibit random spikes in the positive and negative directions. It is
perhaps worth emphasizing that the fluctuations, even for these
system-averaged stress differences, can be large relative to the
locally averaged step sizes, which cannot be distinguished from zero
on these plots.  Figure~\ref{fig:dpdtauhist} shows the probability
distribution functions (PDF's) of $\delta P$ and $\delta \tau$, after
normalizing by their respective standard deviations. The two PDF's for
$\delta p_n$ and $\delta \tau_n$ are virtually indistinguishable, and
decay exponentially for positive and negative values, with a somewhat
more rapid decay for positive values. Sharp drops of $P$ and $\tau$
occur at steps where the shear is reversed. Each drop is associated
with the release of stored mechanical energy. Interesting issues
include the nature of the energy dissipated, and the relation to the
evolution of the force network (\cite{tordZhangBehr, tord07,
  TordMuthus08}).  For instance, recently, we have found that buckling
of the force chains \cite{tordZhangBehr} is an important mechanism
leading to the loss of energy in shear.  We will address these
questions in future work.

To further examine the fluctuations of $P$ and $\tau$, we analyzed
their power spectra from the curves in Figs.~\ref{fig:pressure_step}
and \ref{fig:shearstress_step}. Here, we use step number, $N_s$ as a
time-like variable, and $(N_s/1000)^{-1}$ as a frequency-like
variable.  The spectra vs. frequency variable, shown in
Fig.~\ref{fig:ptauspectrum} are similar for $P$ and $\tau$, and are
typically broad-band.  They suggest a power law decay with an exponent
close to $-2$ for both $P$ and $\tau$.  Similar behavior for the
high-frequency part of the spectrum has been reported in previous
experiments on continuously sheared 3D granular systems
\cite{miller96}.

\begin{figure}
\centerline{\includegraphics[width=2.5in]{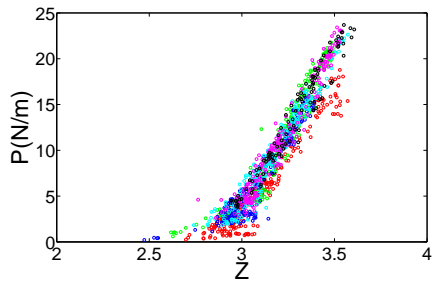}}
\caption{\label{fig:pressureZ}Pressure, $P$, vs. average contact
  number, $Z$. Different colors correspond to data from different
  shear cycles. Data points from the first shear cycle, shown in red,
  deviate slightly from other shear cycles.}
\end{figure}

\begin{figure}
\centerline{\includegraphics[width=2.5in]{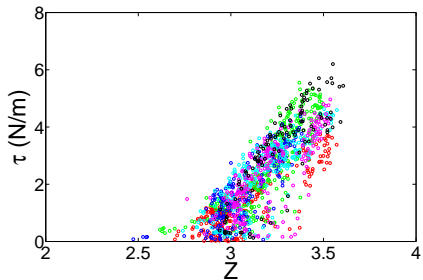}}
\caption{\label{fig:tauZ}Shear stress, $\tau$ vs. average contact
  number, $Z$.}
\end{figure}

\begin{figure}
\centerline{\includegraphics[width=2.5in]{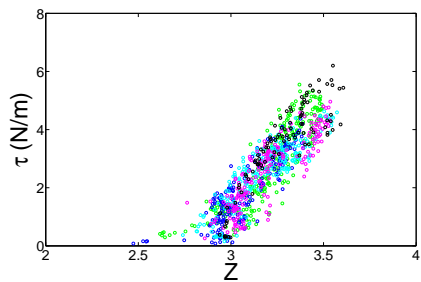}}
\caption{\label{fig:tauZnu}Shear stress, $\tau$ vs. average contact
  number, $Z$ after removal of data points from the first shear cycle
  and data points where the strong network direction, as measured by
  $\vartheta$, is switching directions, as in
  Fig.~\ref{fig:contactAngle_finer}.}
\end{figure}

It is clear from Figs.~\ref{fig:z_half}, \ref{fig:p_half}, and
\ref{fig:tau_half} that the stresses and $Z$ are hysteretic in the
strain, i.e., that the strain does not provide a unique
characterization of a state. It is then interesting to ask whether
there is some other quantity that better characterizes the nature of a
given state, and in particular whether there is a relation among $Z$
and $P$ and $\tau$.  For jamming of spherical (circular in 2D)
particles under isotropic stress conditions, the key control parameter
is the density/packing fraction. Then, both $Z$ and $P$ are functions
of $\phi$, and consequently, $P$ is a function of $Z$.  Here, however,
the packing fraction is constant.  Yet, starting from an unjammed
state, we arrive at a state which is jammed when we apply sufficient
shear.  In order to address what might control jamming in this case,
we note that Figs.~\ref{fig:z_half}, \ref{fig:p_half}, and
\ref{fig:tau_half} show similar shapes. Drawing on the isotropic
stress case, we ask whether a relation exists between $Z$ and the
stresses.

Indeed, Fig.~\ref{fig:pressureZ} shows that data for $P$ vs. $Z$ fall
on a nearly common curve. In a similar fashion, $\tau$ vs. $Z$ falls
on a nearly common curve.  For $\tau$ vs. $Z$, the relative scatter is
higher.  However, there is a systematic part of the $\tau$ vs. $Z$
data that fall below the weight of the curve.  These data correspond
to relatively small ranges of strain following a reversal.  As
$\vartheta$ switches direction, the system passes through a more
nearly isotropic state.  In addition, $\tau$ vs. $Z$ data from the
first cycle start from an isotropic state for which $\tau= 0$.  For
the first part of that cycle, the system retains some memory of its
initial state. If the data from the first cycle and immediately after
reversals are removed, the results for $\tau$ vs. $Z$ yield a collapse
that is comparable to that for $P$ vs. $Z$ as seen in
Fig.~\ref{fig:tauZnu}.  Although $P$ and $Z$ do collapse, the spread
of the data points around the curve is still quite big. This spread
reflects statistical fluctuations, which as shown above, can be large
even for the system-averaged $P$ and $\tau$.  An interesting
observation is that except for the switching regimes, in the mean, the
ratio $\tau/P = constant$.  In this case, the relation follows because
$\tau$ and $P$ are separately (essentially) linear functions of $Z$
and both vanish at the common value $Z \simeq 3$ where the system
first jams.

\begin{figure}
\centering
\includegraphics[width=2.5in]{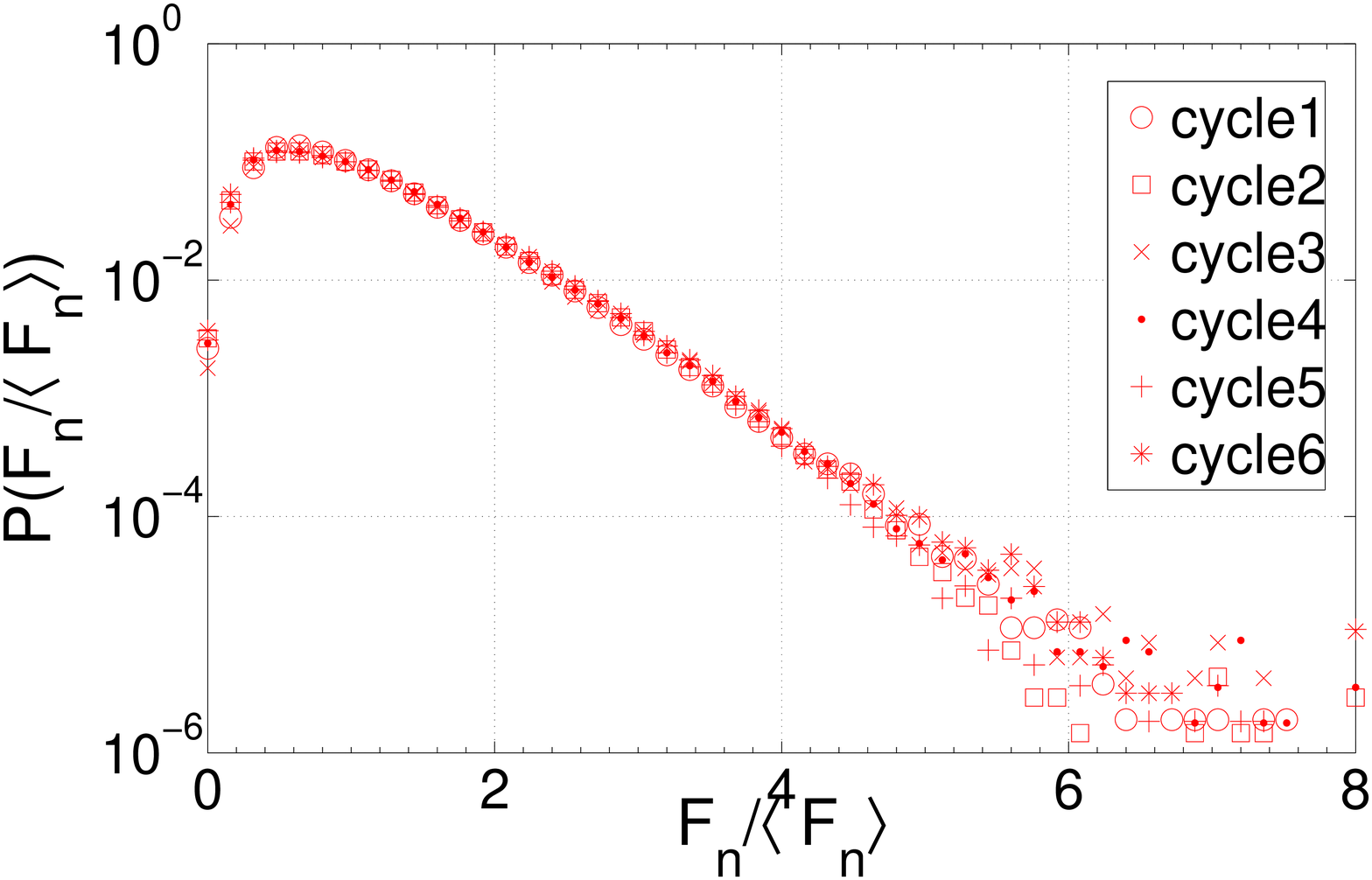}
\caption{\label{fig:Fn_norm_semilog} Data for the distribution of
  normal forces, $F_n$, expressed as $F_n/\langle F_n \rangle$ of all
  six shear cycles. We show data from all strain steps collectively
  for each cycle.  Here, data for $F_n/\langle F_n \rangle$ are
  normalized by $\langle F_n \rangle$ for the given step.}
\end{figure}

\begin{figure}
\centerline{\includegraphics[width=2.5in]{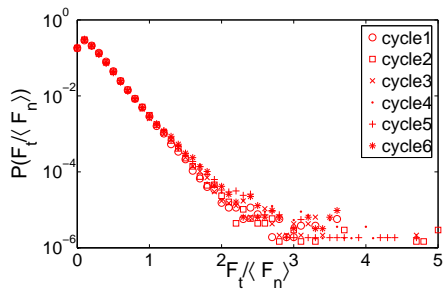}}
\caption{\label{fig:Ft_norm} Data for the distribution of normalized
  tangential forces, $F_t/\langle F_n \rangle$ of all six shear
  cycles.  As for the distribution of normal forces, the statistics
  are combined for all steps within each cycle, where the
  normalization, $\langle F_n \rangle$ is made for each step.  }
\end{figure}

\begin{figure}
\centering
\includegraphics[width=4in]{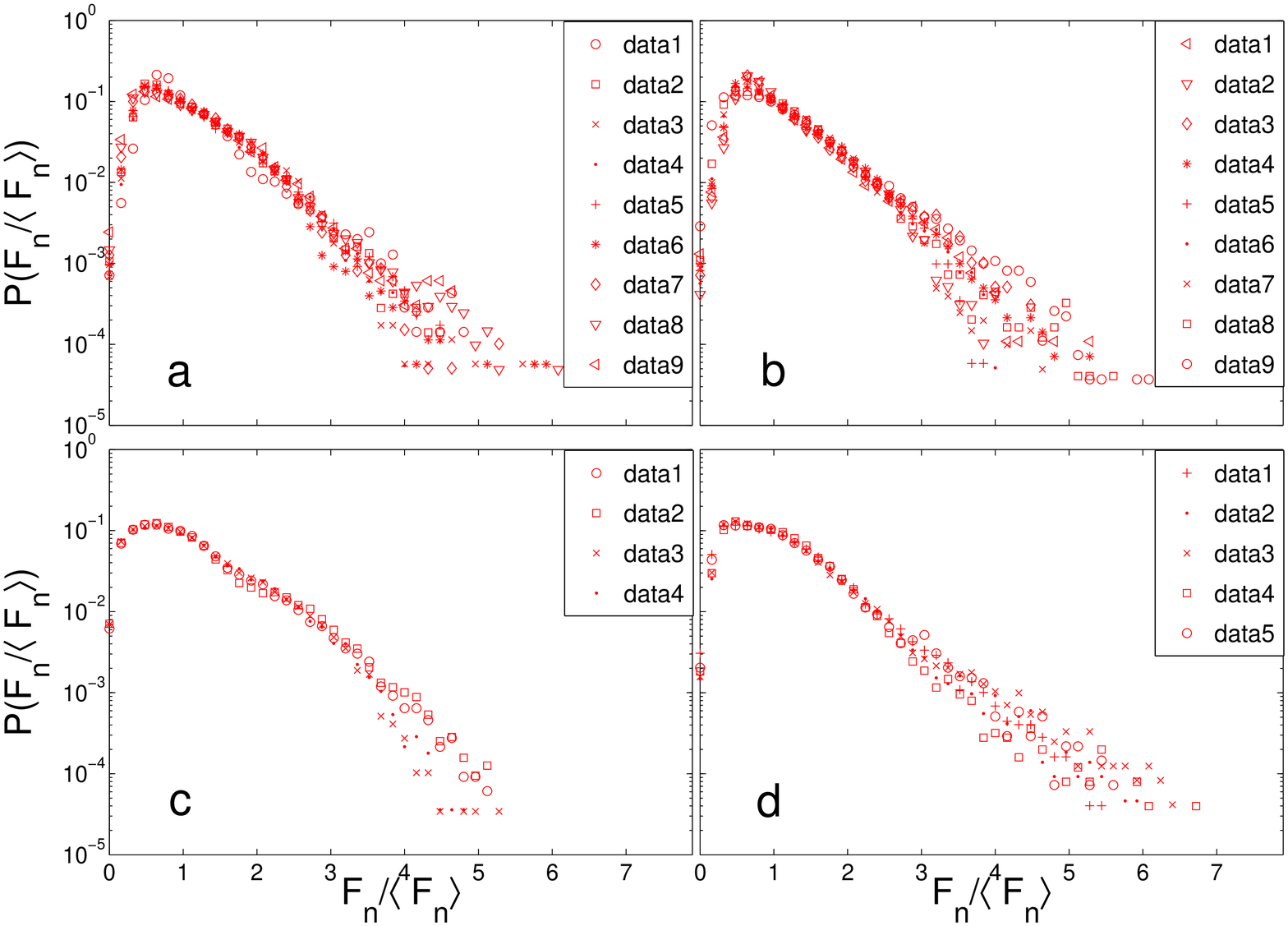}
\caption{\label{fig:cycle1_Fn} Data for the distribution of normal
  forces, $F_n$, expressed as $F_n/\langle F_n \rangle$ in the first
  shear cycle.  (a) distribution $P(F_n/\langle F_n\rangle)$ for
  forward shear, $\epsilon>0$.  (b) $P(F_n/\langle F_n\rangle)$ for
  reverse shear when $\epsilon \geq 0$.  (c) $P(F_n/\langle
  F_n\rangle)$ for reverse shear when $\epsilon<0$.  (d)
  $P(F_n/\langle F_n\rangle)$ for forward shear when $\epsilon \leq
  0$. To improve statistics, each data point on the graph includes a
  set of $F_n/\langle F_n \rangle$ from 10 neighboring steps in most
  cases. A summary of each data point and its corresponding steps and
  strains can be found in Table~\ref{tab:fhist_datai}.  }

\end{figure}

\begin{figure}
\centering
\includegraphics[width=4in]{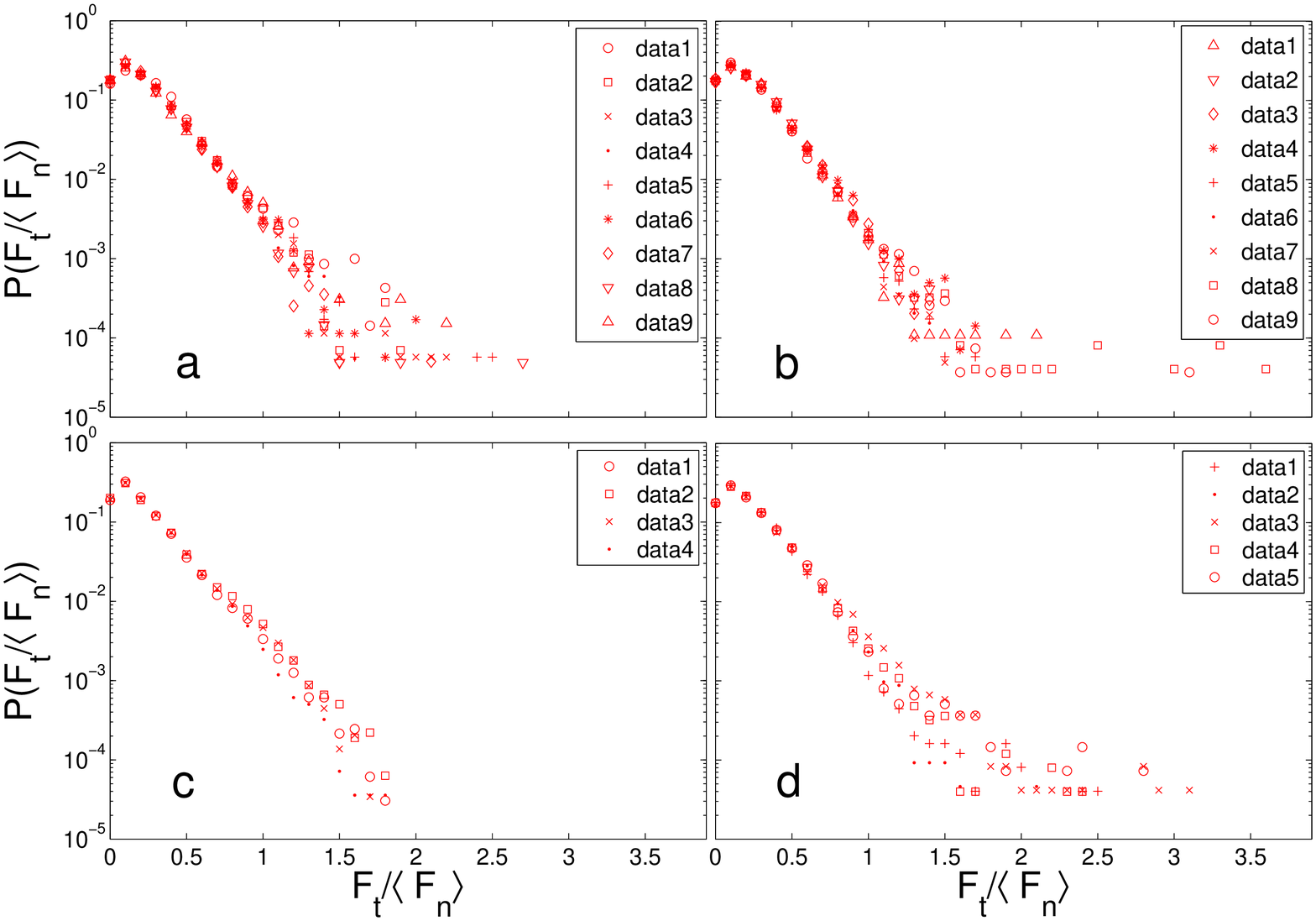}
\caption{\label{fig:cycle1_Ft} Data for the distribution of tangential
  forces, $F_t$, expressed as $F_t/\langle F_n \rangle$ in the first
  shear cycle.  (a) distribution $P(F_t/\langle F_n\rangle)$ of the
  forward shear when the strain $\epsilon>0$.  (b) $P(F_t/\langle
  F_n\rangle)$ of the inverse shear when $\epsilon \geq 0$.  (c)
  $P(F_t/\langle F_n\rangle)$ of the inverse shear when $\epsilon<0$.
  (d) $P(F_t/\langle F_n\rangle)$ of the forward shear when $\epsilon
  \leq 0$. To improve statistics, each data point on the graph
  includes a set of $F_t/\langle F_n \rangle$ from $\sim$10 neighboring
  steps. See details in Table~\ref{tab:fhist_datai}.}
\end{figure}

Additional statistical measures include the distribution of contact
forces, including the normal force distribution given in
Fig.~\ref{fig:Fn_norm_semilog} and the tangential force distribution
given in Fig.~\ref{fig:Ft_norm}. In these figures, we have organized
the data by cycle number, and we have combined data for different
steps within a cycle.  Here, we normalized the data for $F_n $ or
$F_t$ for each step by the the mean normal force $\langle F_n \rangle$
at that step.  To justify that this is legitimate, we plot the
distributions of $F_n/\langle F_n \rangle$ and $F_t/\langle
F_n\rangle$ for different strain steps within the first shear cycle in
Figs.~\ref{fig:cycle1_Fn} and \ref{fig:cycle1_Ft}.  Several
neighboring steps, ten for most data points and three to five for data
near a reversal, are combined for each data point on the plots. The
details are summarized in Table~\ref{tab:fhist_datai}. Some strain
steps may have a longer tail than others but their general shapes are
more or less similar, and in particular, there is no systematic
difference between distributions for different steps. As always, the
tails show bigger scatter due to limited statistics.

The distributions of normal forces show a common form consisting of a
nearly exponential fall-off at large $F_n/\langle F_n \rangle$ and a
peak at low $F_n/\langle F_n \rangle$.  At the extreme tails, the
distributions differ somewhat, but this is to be expected because the
statistics are limited there. $P(F_t/\langle F_n\rangle)$ also shows
an exponential decay, except that the tangential force distribution,
as a function of $P(F_t/\langle F_n\rangle)$, decays faster than that
for the normal forces. However, this is simply due to the choice of
normalization.  That is, if the tangential forces were normalized by
$\langle F_t \rangle$, then the rate of exponential fall-off would be
comparable to that for the normal force distributions.  For small
$F_n/\langle F_n \rangle$ and $F_t/\langle F_n\rangle$, both
distributions fall below exponentials. Although some of the fall-off
is due to the experimental lower limit of force detection, we believe
that this is a relatively minor effect.  Specifically, the shape of
the distribution is not particularly sensitive to the mean force.

\begin{table}
\caption{\label{tab:fhist_datai} Summary of step numbers and strains
  for each group labeled data$_i$ in Figs.~\ref{fig:cycle1_Fn},
  \ref{fig:cycle1_Ft}.}
\begin{tabular}{|l|l|l|l|}
\hline
Panel & data$_i$ & steps & $\epsilon$ \\
\hline
a & 1 & 2-11 & 0.0067 $\sim$ 0.0367 \\
  & 2 & 12-21 & 0.04 $\sim$ 0.07 \\
  & 3 & 22-31 & 0.0733 $\sim$ 0.1033 \\
  & 4 & 32-41 & 0.1067 $\sim$ 0.1367 \\
  & 5 & 42-51 & 0.14 $\sim$ 0.17 \\
  & 6 & 52-63 & 0.1733 $\sim$ 0.21 \\
  & 7 & 64-73 & 0.2133 $\sim$ 0.2433 \\
  & 8 & 74-83 & 0.2467 $\sim$ 0.2767 \\
  & 9 & 84-86 & 0.28 $\sim$ 0.2867 \\
\hline
b & 9 & 88-97 & 0.28 $\sim$ 0.25 \\
  & 8 & 98-107 & 0.2467 $\sim$ 0.2167 \\
  & 7 & 108-117 & 0.2133 $\sim$ 0.1833 \\
  & 6 & 118-127 & 0.18 $\sim$ 0.15 \\
  & 5 & 128-137 & 0.1467 $\sim$ 0.1167 \\
  & 4 & 138-147 & 0.1133 $\sim$ 0.0833 \\
  & 3 & 148-157 & 0.08 $\sim$ 0.05 \\
  & 2 & 158-167 & 0.0467 $\sim$ 0.0167 \\
  & 1 & 168-177 & 0.0133 $\sim$ -0.0167 \\
\hline
c & 1 & 178-187 & -0.02 $\sim$ -0.05 \\
  & 2 & 188-197 & -0.0533 $\sim$ -0.0833 \\
  & 3 & 198-207 & -0.0867 $\sim$ -0.1167 \\
  & 4 & 208-217 & -0.12 $\sim$ -0.15 \\
\hline
d & 5 & 218-227 & -0.1467 $\sim$ -0.1167 \\
  & 4 & 228-237 & -0.1133 $\sim$ -0.0833 \\
  & 3 & 238-247 & -0.08 $\sim$ -0.05 \\
  & 2 & 248-257 & -0.0467 $\sim$ -0.0167 \\
  & 1 & 258-302 & -0.0133 $\sim$ 0 \\
\hline
\end{tabular}
\end{table}

\section{Connection to shear localization and force chain evolution}

We next explore possible connections between the hysteresis observed
in these experiments and two defining aspects of material behavior
under shear: shear banding and stick-slip
(e.g. \cite{alonsomarroquin_06, tord07}).  These two mechanisms are
related and are both governed by force chain/force network evolution.
Note that we use the term `stick-slip' to signify fluctuations in the
macroscopic stress, in particular, that of the stress ratio, $\tau/P$.
Specifically, stick and slip events are periods where the stress ratio
increases and decreases, respectively, with increasing strain.  A slip
event is most often due to the collapse of force chains by buckling
and hence is accompanied by the release of stored energy, accumulated
in force chains during the preceding stick event.

\begin{figure}
\centerline{\includegraphics[width=4in]{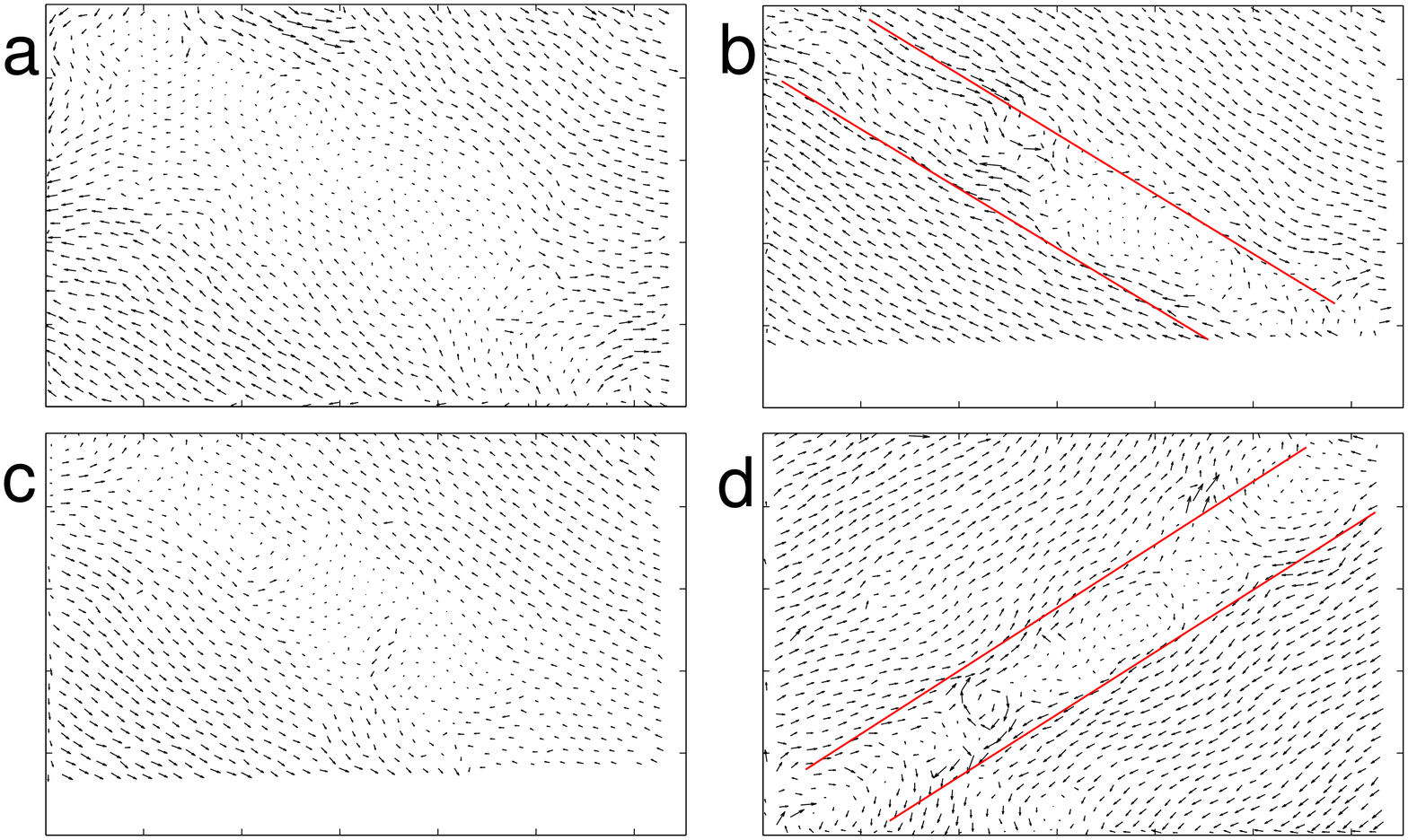}}
\caption{\label{fig:shearbanddisp} Displacement fields showing shear
  bands in forward and reverse shear. These four fields correspond to
  the small particle displacement right after applying small
  deformations to the four images shown in
  Fig.~\ref{fig:force-chains}. The red lines drawn in (b) (d) are a
  guide to the eye, indicating the regions of shear localization.
  Note that in parts (c) and (d) one or more side walls occupies part
  of the image.}
\end{figure}

\begin{figure}
\centerline{\includegraphics[width=2.5in]{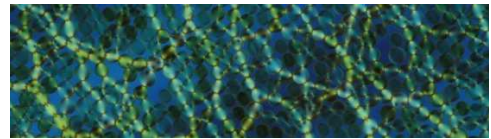}}
\caption{\label{fig:shearbandfc} Force chains in the shear band
  regime. This image is a slice of the main diagonal regime of image
  (b) in Fig.~\ref{fig:force-chains}. Here the image is rotated to
  make the original diagonal line horizontal.}
\end{figure}

In what follows, we examine the evolution of the force network and its
effect on the macroscopic stress.  We confine our attention to the
different time steps for the forward shear. These data are from a run
with a slightly higher density, $\phi=0.76$.  As shown earlier, the
force network evolution in the forward and reverse shear cycles
exhibit fairly universal statistics.  To proceed, we employ two
algorithms.  The first identifies force chain particles via a
so-called particle load vector.  For each particle, the local force
moment tensor, $\hat{\sigma}_{ij}$, as defined earlier, is computed.
The largest eigenvalue of this tensor and its associated eigenvector
are then used to define, respectively, the magnitude and direction of
the particle load vector. The direction of force transmission is
dictated by the direction of the particle load vector. Groups of
particles whose particle load vectors line up
within a predefined narrow angular range, and whose particle load
magnitude is above the global average value, constitute a force
chain. This procedure has been incorporated into an algorithm that
takes contact force data as the known input, and provides the force
chain particles, and hence the force chain particle network, as the
output. Complete details of this procedure and its associated
algorithm are provided elsewhere
\cite{muthuswamy_jstat_06,peters_pre_05}.

The second algorithm was developed for the purposes of identifying
parts of the force chain particle network that have undergone
buckling, i.e. {\it buckled force chain ``BFC'' segments}
\cite{tord07}. A strain interval of interest,
$[\varepsilon^A,\varepsilon^B]$ is chosen: for example, that which
spans a drop in stress ratio or an ``unjamming event'', or a single
time step in a DEM simulation. A set of three filters is then applied:
(a) eliminate all particles not in force chains at $\varepsilon^A$;
(b) out of those remaining, eliminate those which have not decreased
in potential energy; (c) out of those remaining, identify and isolate
all 3-particle segments which have buckled. To determine if a segment
has buckled, we consider the angle between the branch vectors from the
central particle to the two outer particles. The decrease in this
angle over the interval in question is defined as being twice the
buckling angle, $\theta_b$.  Then, a buckling segment is simply one
where $\theta_b>0$.  The set of particles remaining after all three
filters have been applied is the set referred to hereafter as BFCs.
Later, we consider populations of BFCs in distinct subsets, where each
subset is distinguished by a predefined nonzero buckling threshold
$\theta_{b}^{*}$ that member BFCs must satisfy over the given strain
interval.  Complete details of this entire procedure and associated
algorithm are provided elsewhere \cite{tord07}.

The specimen deforms in the presence of a shear band, as shown in
Fig.~\ref{fig:shearbanddisp}. The band is backward inclined in the
forward shear and forward inclined in the reverse shear.  As the
material is sheared in a given direction, two triangular blocks slide
over the shear band in opposing directions, effectively leading to
simple shear across the shear band \cite{thornton_pm_06}.

Although the material is dilated in the shear band, the force chains
pass right through these bands, without substantial changes, due to
force balance (Fig.~\ref{fig:shearbandfc}). During `stick events',
regimes of strain where the force network is stable, the primary force
chains provide the major resistance to motion in the compressive
direction. Weak secondary force chains still exist in the dilation
direction and serve to `prop up' the primary force chains.  Primary
force chains, laterally confined by weak network neighbors, are
subject to axial compression and often fail via buckling.  Secondary
force chains, being in the direction of shear, tend to fail by
extension.

\begin{figure}
\centerline{\includegraphics[width=2.5in]{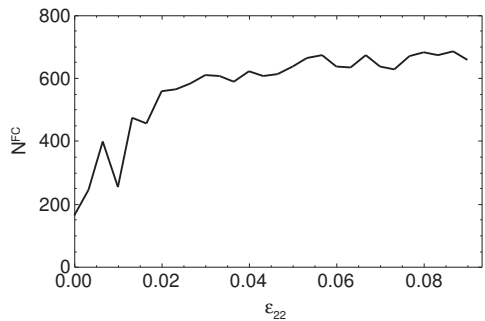}}
\caption{\label{fig:fcpop}Evolution with axial strain,
  $\epsilon_{22}$, of the population of force chain particles,
  $N^{FC}$.}
\end{figure}

\begin{figure}
\centerline{\includegraphics[width=2.5in]{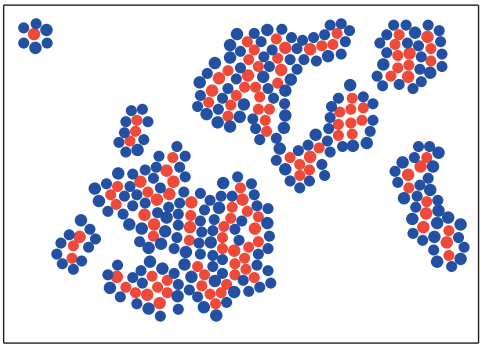}}
\caption{\label{fig:cbfcs}Spatial distribution of force chain
  particles undergoing buckling (red) across the strain interval
  $\epsilon_{22}=0.060$ to $0.077$, together with their confining
  neighbors (blue). A buckling threshold of
  $\theta_{b}^{*}=2^{\circ}$ is used.  Data shown here correspond to
  the central portion of the sample in the early stages of shear band
  development.  We note that the width of the shear band narrows for
  higher strains.}
\end{figure}

\begin{figure}
\centerline{\includegraphics[width=2.5in]{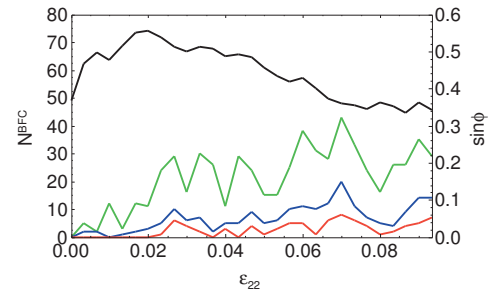}}
\caption{\label{fig:bfcpop}Evolution with axial strain,
  $\epsilon_{22}$, of population of buckled force chain segments,
  $N^{BFC}$, for various buckling thresholds, $\theta_{b}^{*}$. Green,
  blue and red lines correspond to $\theta_{b}^{*}=1^{\circ}$,
  $2^{\circ}$ and $3^{\circ}$, respectively. Also shown is stress
  ratio (black).}
\end{figure}

\begin{figure}
\centerline{\includegraphics[width=2.5in]{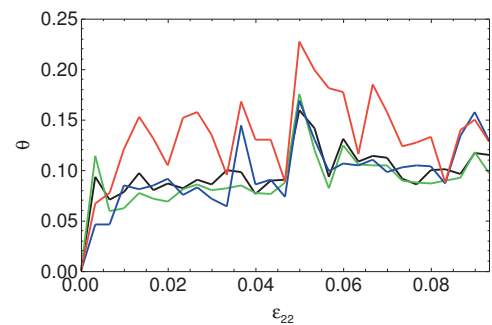}}
\caption{\label{fig:relrot}Evolution with axial strain,
  $\epsilon_{22}$, of the relative rotation at contacts per step,
  $\theta$, for various types of contacts. Black, green, blue and red
  correspond to FC (force chain) to FC, FC to WN (weak network), WN to
  WN and BFC (buckling force chain) to BFC contacts, respectively. The
  unit of $\theta$ is radians.}
\end{figure}

As mentioned earlier (see discussion around
Figs.~\ref{fig:pressure_step} and \ref{fig:shearstress_step}), there
is evidence of local failure events in the force network throughout
the loading, both in the forward and reverse shear.  These failure
events are due to the continual collapse of old and formation of new
force chains.  The failures are concentrated mainly in the shear band
where the mode of deformation of the material is essentially one of
simple shear \cite{thornton_pm_06}.  To unravel the mechanisms behind
the hysteresis in the normal and shear components of stress, we
examine the contribution to these stresses from force chains versus
weak network particles. Recall that we confine our attention to the
forward shear in the first shear cycle.  In Fig.~\ref{fig:fcpop}, we
show the strain evolution of the population of force chain particles.
As expected, there is an initial increase in the population as load is
increased, followed by a near constant value for large strains.  Note
here, that the threshold value for the particle load vector (i.e. the
global average) of each particle classified as a force chain is
increasing with strain.

In light of Figs.~\ref{fig:shearbanddisp}b and
\ref{fig:shearbanddisp}d, the relative rotations between force chains
are particularly relevant here.  Discrete element simulations and
photoelastic disk experiments of Oda and co-workers
(e.g. \cite{oda_geo_80, oda_geo_98, IwashitaOda00, oda_geo_04}) and by
Veje et al. and Utter et al.\cite{veje99,utter04a} have shown that
particle rotations concentrate in the shear band.  More recent studies
confirm this, but additionally found that rotational motions dominate
during slip events inside the shear band (\cite{tordZhangBehr, tord07,
  TordMuthus08}).  These slip events are governed by the failure of
force chains by buckling.  The location and deformation periods during
which relatively large particle rotations occur do indeed coincide
with the location and incidences of force chain buckling: see
Fig.~\ref{fig:cbfcs} and Fig.~\ref{fig:bfcpop}.  Rotation is a key
mechanism in force chain buckling as is evident in
Fig.~\ref{fig:relrot} which shows that the greatest relative rotations
are sustained at contacts between particles in buckling force chains.

\section{CONCLUSIONS}

In this work, we have explored the evolution of force and contact
networks for cyclic shear of a dense granular material.  Starting from
an initially unjammed low-density state, jamming occurs for
sufficiently large shear.  This is, in fact, associated with the
phenomena of Reynolds dilatancy.  In particular, during the first
shear cycle, the system reaches $Z = 3$ around step 70 corresponding
to a shear strain of $\epsilon=23.3\%$ . During much of each cycle,
the system is above a jamming threshold for which the mean contact
number is $Z \simeq 3$.  Although the stress components, $Z$, etc. are
strongly hysteretic in strain, we empirically find that $P$ is
strongly correlated with the mean contact number when the system is
jammed.  The same is true for $\tau$ vs. $Z$ except immediately after
a strain reversal, when $\tau$ can become small.  Again, excluding
regions immediately following reversals, the mean values for $P(Z)$
and $\tau$ approach zero linearly, within our resolution, as $Z
\rightarrow 3$ from above.  In this case, the ratio $\tau/P$ is, on
average, constant.

We have applied a correction to the data to account for contacts which
are below the experimental detection threshold. This correction was
most important for very low stress states. For $Z$, the correction
could reach $\sim 15\%$, although it was also applied to the pressure
and the shear stress, $\tau$.  The correction was implemented by
assuming that the force distributions have (close to) universal
exponential forms.  A more accurate correction technique will be
implemented in future work, but the present approximation is
reasonable, given the statistical scatter of the data.  We note that
the jammed states are generally inhomogeneous in the density/packing
fraction, since these are characterized by shear bands where the
material is locally dilated and where much of the motion occurs.

We then analyzed the evolving force network in the forward shear using
algorithms that distinguish contributions to macroscopic stress of
particles from the strong and weak contact force network.  Relatively
large rotations develop during the buckling of force chains.  These
buckling events, which are present throughout the loading history, are
primarily confined to the shear band and dominate during slip events
or periods where stress ratio decreases with increasing strain.  A
detailed analysis of such internal failure events by buckling and
their connection to structural evolution of the force and contact
networks in cyclic shear loading is the subject of an ongoing
investigation.

{\bf Acknowledgments} RPB acknowledges support from ARO grant
W911NF-07-1-0131 and NSF grant DMR-0555431; AT acknowledges the
support of the Australian Research Council (DP0772409) and the US Army
Research Office (W911NF-07-1-0370).

\bibliographystyle{apsrev}
\bibliography{granmat12-26ANT}

\end{document}